\documentclass[numberedappendix]{emulateapj}
\newcommand{\ack}{\acknowledgements}
\usepackage{apjfonts}

\usepackage{natbib}
\usepackage{graphicx}
\usepackage{amssymb}

\defcitealias{tchekhovskoy_ff_jets_2008}{TMN08}

\begin{document}
\def\rsing{R_{\rm sing}}

\title{STABILITY OF RELATIVISTIC FORCE-FREE JETS}

\author{Ramesh Narayan$^1$,
Jason Li$^2$ and
Alexander Tchekhovskoy$^3$}
\altaffiltext{1}{Harvard-Smithsonian Center for Astrophysics, 60 Garden 
Street, Cambridge, MA 02138, rnarayan@cfa.harvard.edu}
\altaffiltext{2}{Department of Astrophysical Sciences, Princeton University,
Peyton Hall, Ivy Lane, Princeton, NJ 08544, jgli@princeton.edu}
\altaffiltext{3}{Harvard-Smithsonian Center for Astrophysics, 60 Garden 
Street, Cambridge, MA 02138, atchekho@cfa.harvard.edu}
%\ead{rnarayan@cfa.harvard.edu}

\def\vB{\vec{B}}
\def\vE{\vec{E}}
\def\vj{\vec{j}}
\def\vnab{\vec\nabla}
\def\pa{\partial}
\def\Rfp{R_{\rm fp}}
\def\RLC{R_{\rm A}}

\begin{abstract}
We consider a two-parameter family of cylindrical force-free equilibria, modeled to match numerical simulations of relativistic force-free jets.  We study the linear stability of these equilibria, assuming a rigid impenetrable wall at the outer cylindrical radius $R_j$.  We find that equilibria in which the Lorentz factor $\gamma(R)$ increases monotonically with increasing radius $R$ are stable.  On the other hand, equilibria in which $\gamma(R)$ reaches a maximum value at an intermediate radius and then declines to a smaller value $\gamma_j$ at $R_j$ are unstable.  The most rapidly growing mode is an $m=1$ kink instability which has a growth rate $\sim(0.4/\gamma_j)(c/R_j)$.  The $e$-folding length of the equivalent convected instability is $\sim2.5\gamma_jR_j$.  For a typical jet with an opening angle $\theta_j \sim{\rm few}/\gamma_j$, the mode amplitude grows weakly with increasing distance from the base of the jet, much slower than one might expect from a naive application of the Kruskal-Shafranov stability criterion.\end{abstract}

%\noindent{\it Keywords\/}:
\maketitle

\section{Introduction}

Relativistic jets in astrophysical sources have been known for many
decades.  Although their enormous power, large Lorentz factor and
strong collimation have been widely studied, these phenomena still
lack an accepted explanation.  An even greater mystery is the
remarkable coherence and apparent stability of jets over very large
length scales.  This is the topic of the present paper.

The most promising models of relativistic jets involve acceleration
and collimation by magnetic fields with footpoints attached to a
spinning black hole or neutron star or accretion disk.  The forced
rotation of the field lines induces a strong toroidal component of the
magnetic field, which is responsible for accelerating the jet
\citep*[e.g.,][hereafter \citetalias{tchekhovskoy_ff_jets_2008}; and
references therein]{nar07,tchekhovskoy_ff_jets_2008}.  In this picture
the toroidal component of the field dominates over other field
components.

According to the well-known Kruskal-Shafranov criterion
\citep[e.g.,][]{bat78}, cylindrical magnetohydrodynamic (MHD) 
configurations in which the
toroidal field dominates are violently unstable to the $m=1$ kink
instability (also called the screw instability).  The KS criterion for
instability is
\begin{equation}
\left|\frac{B_\phi}{B_p}\right| > \frac{2\pi R_j}{z},
\label{KScriterion}
\end{equation}
where $B_\phi$ and $B_p$ are the toroidal and poloidal magnetic field
strengths, $R_j$ is the cylindrical radius of the jet and $z$ is the
length of the jet (i.e., distance from the base of the jet).  Typical
jet models, including the ones described in this paper (see \S2.2),
have $B_\phi\sim\gamma_j B_p$, where $\gamma_j\gg1$ is the Lorentz
factor of the jet, and they have jet angles $\theta_j\sim R_j/z\sim
{\rm few}/\gamma_j$.  Substituting these scalings in equation
(\ref{KScriterion}), it is obvious that the KS instability criterion
is easily satisfied in relativistic jets.  We therefore expect
astrophysical jets to be violently unstable, as argued for example by
\citet{begelman_pinch_instability_1998} and \citet{li00}.  Yet, jets
in nature are apparently quite stable.  How is this possible?

Many authors have investigated this question.  They have used jet
models with a variety of velocity profiles, geometrical shapes,
composition and boundary conditions
\citep{kad66,bat78,ftz78,ben81,pc85}, and applied both analytical and
numerical methods \citep{ip96,begelman_pinch_instability_1998,lyu99,
li00,lba00,alb00,tom01,mhn07,mso08,mb08}.  As a result of this large
body of work, several kinds of unstable modes have been identified:
reflection modes, Kelvin-Helmholtz modes, current-carrying modes, etc.
Unfortunately, it is difficult to synthesize the results and extract
universal principles.

A fruitful approach in this field is to reduce relativistic jet models
to their barest minimum.  One such approach is to consider {\it
force-free} jet models in which one ignores the inertia and pressure
of the plasma and considers only charges, currents and fields.  The
force-free approximation is valid whenever the energy density in
fields dominates over matter energy density, as in pulsar
magnetospheres \citep{gol69,rs75}.  The force-free approximation is
valid also in relativistic MHD jets, at least inside the fast surface
\citep[e.g.,][]{tchekhovskoy_monopole09}.

Theoretical studies of force-free jets have led to the identification
of two distinct stability criteria.  In a detailed analysis,
\citet{ip96} showed that cylindrical force-free jets in which $B_z$ is
independent of $R$ are stable.  \citet{lyu99} considered models with
non-constant $B_z$ and showed that force-free jets are unstable if
\begin{equation}
\frac{dB_z}{dR} < 0,
\label{IPLcriterion}
\end{equation}
i.e., if the poloidal field decreases with increasing distance from
the axis.  We refer to equation (\ref{IPLcriterion}) as the IPL
criterion for instability.  On the other hand, \citet*{tom01} showed
via an approximate analysis\footnote{They effectively restricted their
analysis to the region of the jet inside the light cylinder.
Therefore, the flow speeds they considered were only
quasi-relativistic at best.} that force-free jets are unstable if
\begin{equation}
\left|\frac{B_\phi}{B_p}\right| > \frac{\Omega R}{c},
\label{TMTcriterion}
\end{equation}
where $\Omega$ is the angular velocity of the field line.  This
criterion --- the TMT criterion --- differs from the KS criterion
(\ref{KScriterion}) in that it explicitly accounts for rotation.  It
is also apparently very different from the IPL criterion.

We describe in this paper a class of force-free cylindrical jet
equilibria which closely match the numerical force-free jet
simulations reported in \citetalias{tchekhovskoy_ff_jets_2008}.  Within the context of force-free jets
from rigidly-rotating stars, we believe that this two-parameter family
of equilibria is generic and fairly complete.  We study the stability
properties of these equilibria and attempt to relate our results to
the KS, IPL and TMT criteria (eqs. \ref{KScriterion},
\ref{IPLcriterion} and \ref{TMTcriterion}).

In \S2 we summarize the numerical simulation results of \citetalias{tchekhovskoy_ff_jets_2008} (\S2.1)
and we describe an analytical force-free jet model which matches the
simulation data very closely (\S2.2).  In \S3 we carry out a linear
stability analysis of these equilibria and show that the linear modes
of the system are obtained by solving an eigenvalue problem involving
two coupled differential equations, with appropriate boundary
conditions.  In \S4 we numerically solve the equations and identify the
unstable modes in the system.  We also derive an approximate estimate
for the growth rate of the instability.  We conclude in \S5 with a
summary and discussion.  We use $(r,\theta,\phi)$ for spherical
coordinates and $(R,\phi,z)$ for cylindrical coordinates.

\section{Force-Free Jet Equilibrium}

\subsection{Structure of Force-Free Jets}
\label{eq_structure_jets}

\citetalias{tchekhovskoy_ff_jets_2008} considered a rigidly-rotating star of unit radius ($r=1$)
surrounded by a differentially-rotating infinitely thin disk extending
from $R=1$ outward.  The star was threaded by a uniform radial
magnetic field $B_r$, and the disk was threaded by a power-law
distribution of vertical field,
\begin{equation}
B_z \propto R^{\nu-2}.
\label{Bznu}
\end{equation}  
Using a relativistic force-free code  \citep{gam03,mck04, mck06jf,migmck07, 
tch_wham07}, \citetalias{tchekhovskoy_ff_jets_2008} numerically evolved the system and obtained the
equilibrium configuration of the magnetic field.

Following \citetalias{tchekhovskoy_ff_jets_2008}, we will call the field lines that emerge from the
star as the ``jet'' and the field lines from the disk as the ``wind.''
The critical field line that emerges from the star-disk boundary
defines the boundary between the jet and the wind.  This boundary
starts off at $\theta=\pi/2$ at the surface of the star ($r=R=1$) but
decreases to smaller values of $\theta$ with increasing $r$ (or $z$).

We are primarily interested in the ``jet'' --- the bundle of field
lines attached to the central star.  Since all of these field lines
rotate at the angular velocity $\Omega$ of the star, the Alfv\'en
surface for these lines takes the form of a cylinder --- the ``light
cylinder'' --- with radius $R_A = c/\Omega$.  Field lines become
strongly toroidal once they are outside the Alfv\'en surface, which is
where most of the collimation and acceleration occurs (\citetalias{tchekhovskoy_ff_jets_2008}).

\citetalias{tchekhovskoy_ff_jets_2008} showed that the structure of the jet is strongly affected by the
radial pressure profile of the region surrounding the jet.
Specifically, if we write the radial variation of the confining
pressure as $p\propto r^{-\alpha}$, the jet properties are determined
by the value of $\alpha$.  In the numerical experiments, the pressure
was caused by a force-free disk wind, and $\alpha$ was determined by
the index $\nu$ defined in equation (\ref{Bznu}) according to
\begin{equation}
\alpha=2(2-\nu).
\end{equation}
At distance $z$ along the axis, the cylindrical radius $R_j$ of the
jet is approximately given by
\begin{equation}
R_j \sim z^{\alpha/4}.
\label{Rjz}
\end{equation}  
For all $\alpha<4$, the jet collimates as it moves away from the star
\citep{lyn2006}.  As a result, at a sufficiently large distance from
the central star ($r\gg1$), the jet is nearly cylindrical in shape.

In the asymptotic nearly-cylindrical region of the jet, force balance
in the $R$ direction is described by the following equation (\citetalias{tchekhovskoy_ff_jets_2008}):
\begin{equation}
{d\over dR}\left({B^2-E^2\over 8\pi}\right) + \left({B_\phi^2-E^2\over
4\pi R}\right) + \left({B_p^2-E^2\over 4\pi R_c}\right) = 0,
\label{balance}
\end{equation}
where $B$ is the total magnetic field strength, $B_\phi$ is the
toroidal field strength, and $B_p$ is the poloidal field strength:
\begin{equation}
B=\sqrt{B_p^2+B_\phi^2}, \qquad B_p = \sqrt{B_R^2+B_z^2}.
\label{BBp}
\end{equation}
The electric field $\vec{E}$ is given by
\begin{equation}
\vec{E} = -(\Omega R/c)\,\hat{\phi}\times\vec{B},
\label{Efield}
\end{equation}
where $\Omega$ is the angular velocity of the field line.  The
electric field has only a poloidal component: $E_p=\Omega R B_p/c$.

Each of the three terms on the left-hand side of equation
(\ref{balance}) represents a force in the $-R$ direction.  The
quantity $(B^2-E^2)/8\pi$ is the pressure of the force-free fluid in
the comoving frame; therefore, the first term describes the inward
force due to the gradient of pressure.  The second term arises from
the toroidal curvature of the field line.  The toroidal magnetic field
$B_\phi$ contributes an inward force due to ``hoop stress,'' while the
poloidal electric field $E$ contributes an outward force.\footnote{By
equation (\ref{Efield}), $\vec{E}$ is directly proportional to
$\Omega$, so the latter term results from rotation and may loosely be
viewed as a sort of ``centrifugal force'' (V. Beskin, private
communication).} The third term in (\ref{balance}) gives analogous
contributions from the poloidal curvature of the field line, where
$R_c$ is the poloidal radius of curvature; once again there is an
inward force due to the poloidal magnetic hoop stress and an outward
force due to the electric field.  Note that the contributions
involving $E$ are important only for relativistic flows.  In standard
non-relativistic MHD, one neglects these terms and keeps only the
terms involving $B$.

\begin{figure*}
\begin{center}
\includegraphics[width=0.7\textwidth]{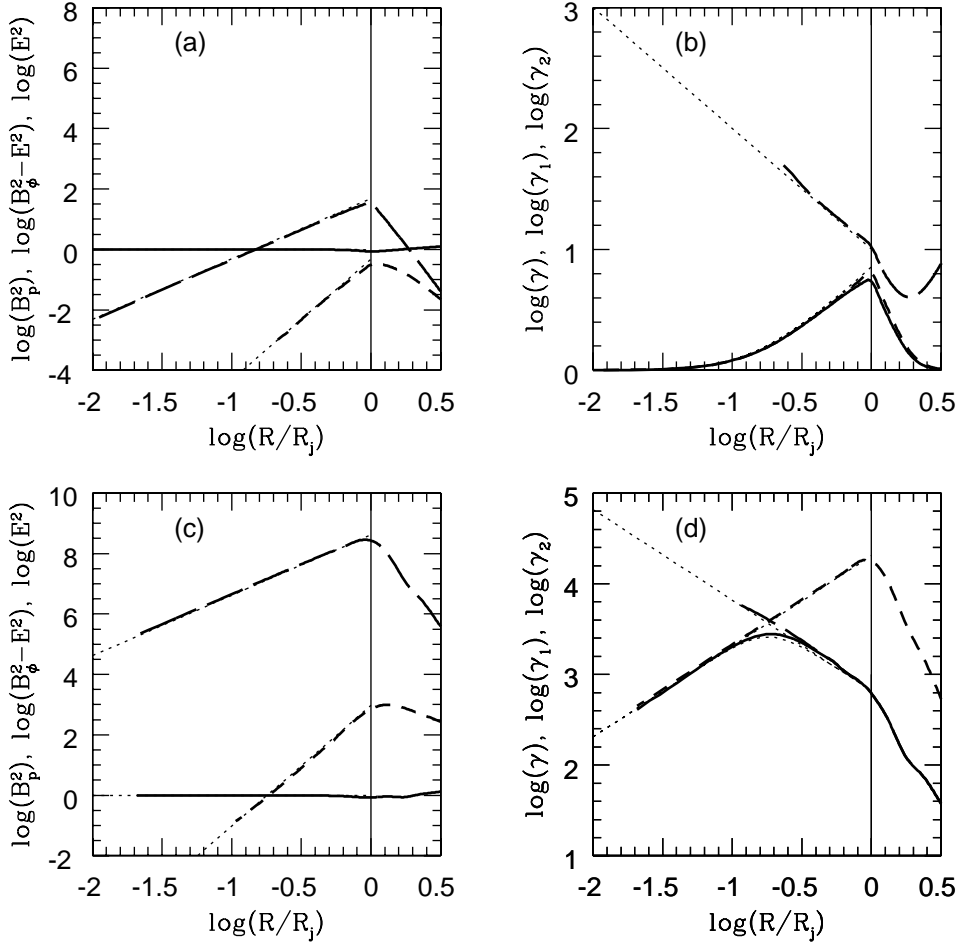}
%\vskip -0.5in
\caption{Numerical results from a force-free jet simulation with
$\alpha=2.5$ (``fiducial model'' corresponding to $\nu=0.75$ in
\citetalias{tchekhovskoy_ff_jets_2008}).  [Panel (a)]: Field components as functions of normalized cylindrical
radius $R/R_j$ at $z=10^2$.  The solid horizontal line shows
$B_p^2(R)/B_p^2(0)$, the short-dashed line shows $[B_\phi^2(R)-E^2(R)]
/B_p^2(0)$ and the long-dashed line shows $E^2(R)/B_p^2(0)$.  The
dotted lines are the corresponding results for the analytical model
with $\gamma_m=6$, $R_m=1.2$ (Model A, \S2.2.2).  The vertical solid
line shows the boundary between the jet and the external confining
medium.  [Panel (b)]: Lorentz factor $\gamma(R)$ (solid line), and the two
approximations, $\gamma_1(R)$ (short-dashed line) and $\gamma_2(R)$
(long-dashed line), at the same $z$.  The dotted lines are from the
analytical model.  [Panels (c), (d)]: Similar to (a), (b), but at $z=10^7$.  The
dotted lines in these panels correspond to the analytical model with
$\gamma_m=2600$, $R_m=0.18$ (Model C, \S2.2.2).}
\end{center}
\label{fig1}
\end{figure*}

\citetalias{tchekhovskoy_ff_jets_2008} showed that models with $\alpha>2$, i.e., $\nu<1$, are good
analogs of relativistic jets found in nature (especially the jets of
gamma-ray burst).  Figure \ref{fig1} shows numerical results corresponding to
the ``fiducial'' force-free simulation in \citetalias{tchekhovskoy_ff_jets_2008} with $\nu=0.75$
(equivalent to $\alpha=2.5$).  Panels (a) and (b) show results for a
relatively near region of the jet at $z=10^2$, and panels (c) and (d)
show results for a more distant region at $z=10^7$.  In each case, the
abscissa corresponds to the cylindrical radius $R$ normalized by the
local ``jet radius'' $R_j$, which is the cylindrical radius of the
last jet field line that separates the jet from the surrounding disk
wind.

The main features of the numerical solution are as follows.  First,
from panels (a) and (c) we see that $B_p$ is essentially constant
inside the jet, showing hardly any variation with $R$.  As we show in
Appendix~\ref{sec_Bp_const}, this is required for force-free jet
solutions that smoothly connect to the central compact
object.\footnote{Asymptotic force-free jet configurations with
non-constant profiles of $B_p$ are certainly possible \citep{ip96},
but there exists no solution that would smoothly connect them to the
compact object.}  Second, since $\Omega$ is constant and $E\propto
\Omega R B_p$ (eq. \ref{Efield}), we have $E^2\propto R^2$.  Third,
$B_\phi$ is almost equal to $E$ and so $B_\phi$ also varies primarily
as $R^2$.  Fourth, $\left|B_\phi\right|$ is slightly larger than $E$ with
$B_\phi^2-E^2 \propto R^4$.  Because of this property, the first two
terms in equation (\ref{balance}) both give an inward force.
Therefore, we obtain the fifth feature of the solution, viz., the
third term in (\ref{balance}), which involves the poloidal curvature
of the field line, is important for force balance.  This is the only
outward force in the balance equation -- it is outward because $E$ is
of order $B_\phi$ and is much greater than $B_p$ outside the light
cylinder.  This force has to balance the other two terms.

The velocity of a force-free flow is usually identified with the
drift velocity,
\begin{equation}
\frac{\vec{v}}{c} = \frac{\vec{E}\times\vec{B}}{B^2},
\label{vdrift}
\end{equation}
and the Lorentz factor is defined correspondingly.  Since $\vec{E}\cdot
\vec{B}=0$, it is easily shown that
\begin{equation}
\gamma^2 = \frac{B^2}{B^2-E^2}.
\label{gammasq}
\end{equation}
Panels (b) and (d) show the variation of $\gamma$ as a function of the
normalized cylindrical radius $R/R_j$ in the numerical model.  \citetalias{tchekhovskoy_ff_jets_2008}
derived two approximate relations for $\gamma$,
\begin{eqnarray}
\gamma_1 &=& [1+(\Omega R/c)^2]^{1/2}, \label{gamma1} \\
\gamma_2 &=& (3R_c/R)^{1/2}, \label{gamma2}
\end{eqnarray} 
and they showed that the net $\gamma$ of the fluid is given to good
accuracy by the following simple formula,
\begin{equation}
{1\over\gamma^2} = 
{1\over\gamma_1^2}+{1\over\gamma_2^2}.
\label{gamma}
\end{equation}
Along each field line, $\gamma$ is initially determined mainly by
rotation, and so $\gamma\approx\gamma_1$.  This is a region of
efficient acceleration which \citetalias{tchekhovskoy_ff_jets_2008} called the {\it first acceleration
regime}.  However, beyond a certain distance from the star, the effect
of poloidal field line curvature becomes important, and $\gamma$
switches to the less efficient $\gamma_2$, the {\it second
acceleration regime}.

Relatively near the star, all field lines in the jet are in the first
acceleration regime and $\gamma(R)$ behaves like $\gamma_1$
(eq. \ref{gamma1}), as seen in panel (b).  Specifically, $\gamma$
increases more or less linearly with $R$ and reaches its maximum value
at the edge of the jet at $R=R_j$.  However, when we consider the jet
at a larger distance from the star, some of the field lines have
already switched to the second acceleration regime, where
$\gamma\sim\gamma_2\propto 1/R$ (see eq. \ref{gamma2}, coupled with
eq. \ref{Rc} below).  We then have the results shown in panel (d),
where the maximum Lorentz factor occurs at some radius $R_m$ inside
the jet, not at the boundary; we have $\gamma\sim\gamma_1\propto R$
for $R \lesssim R_m$ and $\gamma\sim\gamma_2\propto R^{-1}$ for $R_m
\lesssim R<R_j$.  \citetalias{tchekhovskoy_ff_jets_2008} discuss in detail the physics of the two
acceleration regimes.

\subsection{Analytical Jet Model}

The axisymmetric numerical jets models described in \S2.1 have
magnetic and electric field components that are functions of both $R$
and $z$.  This is not convenient for linear perturbation analysis.
Since the numerical models are nearly cylindrical at large distance
(i.e., $dR/dz \ll 1$), we consider now an idealized jet equilibrium
model which is perfectly cylindrical and in which all quantities are
functions only of $R$.  We choose the following specific functional
forms:
\begin{eqnarray}
B_{0R} &=& 0, \label{B0R}\\ 
B_{0\phi} &=& -\left[2(\gamma_m^2-1)(R/R_m)^2 + (R/R_m)^4\right]^{1/2}
\equiv -f(R), \label{B0phi}\\
B_{0z} &=& \exp\left[-3R^2/4(\gamma_m^2-1)R_m^2\right]
\equiv g(R), \label{B0z}\\
E_{0R} &=& -\left[2(\gamma_m^2-1)\right]^{1/2}(R/R_m)
\equiv -h(R), \label{E0R}\\
E_{0\phi} &=& 0, \label{E0phi}\\
E_{0z} &=& 0, \label{E0z}\\
R_c &=& 2(\gamma_m^2-1)R_m^2/3R. \label{Rc}
\end{eqnarray}
The zeros in the subscripts are meant to indicate that all these
quantities refer to the unperturbed model.  The model has two
parameters, $\gamma_m$ and $R_m$, whose meanings are explained below.
For simplicity, we have chosen units such that $B_{0z}=1$ at the jet
axis ($R=0$).  Also, we have assumed that $B_{0z}$ and $\Omega$ are
positive, so both $B_{0\phi}$ and $E_{0R}$ are negative, i.e.,
magnetic field lines are swept backward with respect to the rotation
and the electric field is pointed radially inward.  With this choice
of signs, the three functions, $f(R)$, $g(R)$ and $h(R)$, are
positive.  Note that, in all cases of interest, $g(R)$ is practically
equal to unity.  The particular exponential form given in equation
(\ref{B0z}) is designed to handle small higher-order terms in the
force balance equation (\ref{balance2}), but the deviations of $g(R)$
from unity are tiny and unimportant.\footnote{As a test, in the
stability analysis described later we have done the calculations both
with the full expression for $g(R)$ given in eq. (\ref{B0z}) and with
the simpler choice $g(R)=1$.  The results are practically the same.}

By direct substitution it is easily verified that the above model
satisfies the radial force balance equation (\ref{balance}).  Under
cylindrical symmetry, this equation takes the form:
\begin{equation}
{d\over dR}\left({B_{0\phi}^2+B_{0z}^2-E_{0R}^2\over 8\pi}\right) +
\left({B_{0\phi}^2-E_{0R}^2\over 4\pi R}\right) +
\left({B_{0z}^2-E_{0R}^2\over 4\pi R_c}\right) = 0.
\label{balance2}
\end{equation}
The first two terms are positive, i.e., both represent inward forces,
with the first term providing twice as much force as the second.  The
third term is negative and its magnitude is equal to the sum of the
other two terms.

An important feature of the above model is that the outward force from
the third term involves the poloidal curvature radius $R_c$.
Technically, a perfectly cylindrical model has $R_c\to\infty$.  To get
around this problem, we treat $R_c$ as an externally imposed property
of the solution which is adjusted so as to reproduce the poloidal
curvature force present in the numerical jet model.  In other words,
even though we have straightened out field lines in the $z$-direction
by enforcing cylindrical geometry, we still retain the effect of
poloidal curvature by means of an artificial external force.  This
procedure is analogous to the widely-used shearing sheet approximation
in accretion disk studies \citep{gt78,nar87} in which fluid
streamlines are straightened out in the azimuthal direction, but the
effect of azimuthal curvature is still retained via a Coriolis force.
Note that, apart from the extra term due to poloidal field curvature,
equation (\ref{balance}) is identical to the standard balance
condition derived in other papers in the literature, e.g., equation
(6) in \citet{ip96} or equation (16) in \citet{lyu99}.

To get a better idea of the nature of the above analytical model, we
now make a couple of simplifications.  As already mentioned, $B_{0z}$
is practically independent of $R$ inside the jet.  Also, for highly
relativistic jets, we invariably have $B_\phi^2-E^2 \ll B_\phi^2$.  We
therefore replace equations (\ref{B0phi}), (\ref{B0z}) and (\ref{E0R})
by the following simpler formulae
\begin{eqnarray}
B_{0z} &\approx& 1, \\
B_{0\phi} &\approx& E_{0R} ~=~ -\left[2(\gamma_m^2-1)\right]^{1/2}(R/R_m), \\
B_{0\phi}^2-E^2 &=& (R/R_m)^4.
\end{eqnarray}
By equation (\ref{Efield}), the angular velocity of
rotation of the field lines is given by
\begin{equation}
\Omega = -cE_{0R}/B_{0z}R \approx \sqrt{2(\gamma_m^2-1)}\,(c/R_m),
\label{Omega}
\end{equation}
and is the same for all field lines, as required for a rigidly
rotating star at the base of the jet.\footnote{Since we have designed
our analytical model to match the numerical models of \citetalias{tchekhovskoy_ff_jets_2008}, all of
our models have constant $\Omega(R)$.  It would be straightforward to
generalize the model to non-constant $\Omega(R)$, using additional
parameters.}  Using these simpler expressions, we obtain the following
result for the Lorentz factor:
\begin{eqnarray}
{1\over\gamma^2(R)} &=& {B_0^2-E_0^2\over B_0^2} =
 {B_{0\phi}^2+B_{0z}^2-E_{0R}^2\over B_{0\phi}^2+B_{0z}^2} \nonumber\\
 &\approx& {1+(R/R_m)^4\over 1+2(\gamma_m^2-1)(R/R_m)^2} \label{gammap}\\
 &\approx& {1\over 1+(\Omega R/c)^2} + {R\over3R_c}, \label{gammapp}
\end{eqnarray}
where we have made use of equations (\ref{Rc}) and (\ref{Omega}).  We
thus reproduce the result given earlier in equation (\ref{gamma}).

It is easily shown that $\gamma$ reaches a maximum at $R=R_m$ and
that its value at this radius is equal to $\gamma_m$.  Thus, the two
model parameters $R_m$ and $\gamma_m$ allow us to control the basic
features of the equilibrium.

\subsubsection{$R_m>R_j$: Maximum Lorentz Factor Located at the Jet Boundary}

A jet in which all field lines are in the first acceleration regime
has its maximum Lorentz factor at the boundary of the jet, $R=R_j$.
This corresponds to choosing $R_m>R_j$ in the analytical model, so
that the term $(R/R_m)^4$ in the numerator of equation (\ref{gammap})
can be neglected.  In this case, the profile of $\gamma$ has two
segments: the region of the jet inside the light cylinder
($R<R_A=c/\Omega$) which is not accelerated very much, and the region
outside the light cylinder which has $\gamma$ increasing linearly with
radius,
\begin{eqnarray}
\gamma(R) \approx 
  \cases {
    1,                              &$R<R_A = R_m/\sqrt{2(\gamma_m^2-1)}$, \cr
    R/R_A \approx (R/R_j)\gamma_j,  &$R_A < R < R_j$,
  }
  \nonumber
\end{eqnarray}
where $\gamma_j$ is the Lorentz factor at the jet boundary,
\begin{equation}
\gamma_j \approx \gamma_1(R_j) = \left[1+2(\gamma_m^2-1)(R_j/R_m)^2\right]^{1/2}.
\label{gammaj1}
\end{equation}

\begin{figure}
\includegraphics[width=\columnwidth]{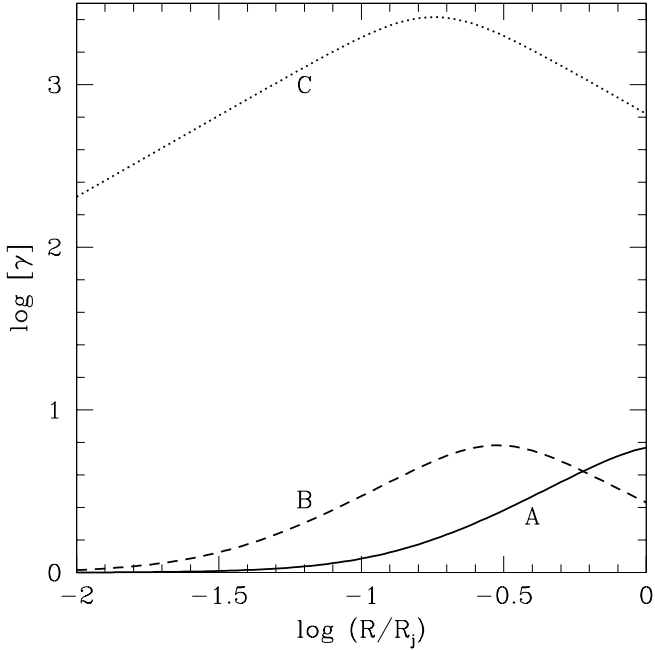}
%\vskip -0.5in
\caption{Profiles of $\gamma$ vs $R/R_j$ for the analytic Models A, B
and C.}
\label{fig2}
\end{figure}

The dotted lines in panels (a) and (b) in Fig. \ref{fig1} show the dependences
of various quantities as a function of $R/R_j$ for a model with
$\gamma_m=6$ and $R_m=1.2R_j$.  The agreement with the numerical
simulation results at $z=10^2$ is striking.  We call the analytical
model with this particular choice of $\gamma_m$ and $R_m$ as Model A:
\begin{equation}
{\rm Model~A}:\qquad\gamma_m=6,\quad R_m=1.2R_j.\label{ModelA}
\end{equation}
The solid line in Fig. \ref{fig2} shows the variation of $\gamma$ as a function
of $R$ for this model.

\subsubsection{$R_m<R_j$: Maximum Lorentz Factor Located Inside the Jet}

As we described in \S2.1, a jet in which some field lines have
switched to the second acceleration regime has its maximum Lorentz
factor inside the jet.  This means $R_m<R_j$.  In this case, the
Lorentz factor $\gamma_j$ at the jet boundary is roughly equal to
\begin{equation}
\gamma_j \approx \gamma_2(R_j) = \left[2(\gamma_m^2-1)\right]^{1/2} R_m/R_j.
\label{gammaj2}
\end{equation}
Now the profile of $\gamma$ has three segments:
\begin{eqnarray}
\gamma(R) \approx
  \cases{
     1,                                       &$R<R_A$, \cr
     R/R_A \approx (R/R_m)\gamma_m,           &$R_A<R<R_m$, \cr 
     (R_m/R)\gamma_m \approx (R_j/R)\gamma_j, &$R_m<R<R_j$.
  }
\end{eqnarray}

The dotted lines in panels (c) and (d) of Fig. \ref{fig1} show model results
corresponding to $\gamma_m=2600$ and $R_m=0.18R_j$.  We find excellent
agreement with the numerical results at $z=10^7$.  We call the
analytical model with these values of $\gamma_m$ and $R_m$ as Model C.
For completeness we also consider a less extreme model called Model B
in which $\gamma_m=6$ and $R_m=0.3R_j$:
\begin{eqnarray}
&~&{\rm Model~B}:\qquad \gamma_m=6,~\qquad R_m=0.3R_j,\label{ModelB} \\
&~&{\rm Model~C}:\qquad \gamma_m=2600,\quad R_m=0.18R_j.\label{ModelC}
\end{eqnarray}
The dashed and dotted lines in Fig. \ref{fig2} show the variations of $\gamma$
as a function of $R$ for these two models.

\section{Linear Perturbation Analysis}

We now consider linear perturbations of the cylindrical equilibrium
described in \S2.2.  The unperturbed state has magnetic and electric
fields
\begin{eqnarray}
\vec{B}_0 &=& B_{0R}\hat{R}+B_{0\phi}\hat{\phi}+B_{0z}\hat{z}, \\
\vec{E}_0 &=& E_{0R}\hat{R}+E_{0\phi}\hat{\phi}+E_{0z}\hat{z},
\end{eqnarray}
where the various components are given by the expressions in equations
(\ref{B0R})--(\ref{E0z}).  As mentioned previously, we choose $B_{0z}$
and $\Omega$ to be positive, so $B_{0\phi}$ and $E_{0R}$ are negative.
The unperturbed current and electric charge are
\begin{eqnarray}
\vec{J}_0 &=& {c\over4\pi}\vec{\nabla}\times\vec{B}_0 = {c\over4\pi}\left[
-\left({dB_{0z}\over dR}+{B_{0z}\over R_c}\right)\hat{\phi}+
{1\over R}{d\over dR}(RB_{0\phi})\hat{z}\right], \nonumber \\
\rho_0 &=& {1\over4\pi}\vec{\nabla}\cdot\vec{E}_0 = {1\over4\pi R}
{d\over dR}(RE_{0R})+{1\over4\pi}{E_{0R}\over R_c}.\nonumber
\end{eqnarray}
Note that we have included terms involving $R_c$ in the unperturbed
current and charge density.  These terms describe the contributions of
poloidal field curvature to the quantities
$\vec{\nabla}\times\vec{B}_0$ and $\vec{\nabla}\cdot\vec{E}_0$,
respectively.  By including these terms, we retain the forces
associated with poloidal field curvature without actually having
curved field lines in the model.

\subsection{The Eigenvalue Problem}

We now consider small perturbations.  Let us write the perturbed
electric field as $\vec{E} = \vec{E}_0 + \vec{E}_1$, where $\vec{E}_1$
is a small perturbation of the form
\begin{equation}
\vec{E}_1=[E_{1R}(R)\hat{R}+E_{1\phi}(R)\hat{\phi}+E_{1z}(R)\hat{z}]
\,\exp(-i\omega t + im\phi+ikz).
\label{eigenfunction}
\end{equation}
Let us similarly write $\vec{B} = \vec{B}_0 +\vec{B}_1$, $\vec{J} =
\vec{J}_0 +\vec{J}_1$, $\rho = \rho_0 +\rho_1$.  Each of these small
perturbations can be expressed in terms of the perturbed electric
field via Maxwell's equations.  From
\begin{equation}
{1\over c}{\partial\vec{B}\over\partial t} = -\vec{\nabla}
\times\vec{E},
\end{equation}
we obtain
\begin{equation}
\vec{B}_1=-{ic\over\omega}\,\vec{\nabla}\times\vec{E}_1.
\end{equation}
From
\begin{equation}
{1\over c}{\partial\vec{E}\over\partial t} = \vec{\nabla}
\times\vec{B} -{4\pi\over c}\vec{J},
\end{equation}
we obtain
\begin{equation}
\vec{J}_1={i\omega\over4\pi}\vec{E}_1 - {ic^2\over4\pi\omega}
\vec{\nabla}\times(\vec{\nabla}\times\vec{E}_1).
\end{equation}
Finally, from
\begin{equation}
\vec{\nabla}\cdot\vec{E}=4\pi\rho,
\end{equation}
we obtain
\begin{equation}
\rho_1={1\over4\pi}\vec{\nabla}\cdot\vec{E}_1.
\end{equation}

Since the perturbed system is force-free, it must satisfy
$\vec{E}\cdot\vec{B}=0$.  The zeroth order terms satisfy this
trivially (as they should).  The first order terms give the condition
$\vec{E_1}\cdot\vec{B_0}+\vec{E_0}\cdot\vec{B_1}=0$.  Substituting for
the various quantities, this condition allows us to solve for
$E_{1\phi}$ in terms of $E_{1z}$:
\begin{equation}
E_{1\phi} = C_1 E_{1z},
\label{Ephi}
\end{equation}
where the function $C_1$ is given by
\begin{equation}
C_1 = {\omega Rg-cmh\over\omega Rf-ckRh}.
\end{equation}
Successive differentiations give
\begin{eqnarray}
E_{1\phi}^{'} &=& C_1 E_{1z}^{'} + C_1^{'} E_{1z},
\label{Ephip} \\
E_{1\phi}^{''} &=& C_1 E_{1z}^{''} + 2 C_1^{'} E_{1z}^{'} + C_1^{''} E_{1z}.
\label{Ephipp}
\end{eqnarray}

We now consider the force balance condition:
$\rho\vec{E}+(1/c)\vec{J}\times\vec{B}=0$.  The zeroth order terms
give
\begin{equation}
\rho_0\vec{E}_0 + {1\over c}\vec{J}_0\times
\vec{B}_0 =0,
\end{equation}
which is simply the equilibrium force balance condition
(\ref{balance2}).  Notice that the poloidal curvature terms in
$\vec{J}_0$ and $\rho_0$ are necessary to satisfy equilibrium in the
unperturbed solution.  From the first order terms in the force balance
equation we obtain
\begin{equation}
\rho_1\vec{E}_0 + \rho_0\vec{E}_1 +{1\over c}\vec{J}_1\times
\vec{B}_0 + {1\over c}\vec{J}_0\times\vec{B}_1 = 0.
\label{forcebalance}
\end{equation}
The $\hat{\phi}$ component of this equation gives a relation between
$E_{1\phi}^{'}$, $E_{1\phi}$, $E_{1z}^{'}$ and $E_{1z}$.  Eliminating
$E_{1\phi}^{'}$ using equation (\ref{Ephip}), we obtain a first-order
differential equation for $E_{1z}(R)$:
\begin{equation}
D_1E_{1z}^{'} + D_2E_{1z} + D_3E_{1R} = 0,
\label{Ezp}
\end{equation}
where $D_1$, $D_2$ and $D_3$ are functions of $R$.  The expressions
are relatively long and we give them in Appendix~\ref{sec_coeffs}.
The $\hat{z}$ component of (\ref{forcebalance}) has no new
information; it just gives back equation (\ref{Ephi}).  The $\hat{R}$
component, however, gives a new relation between the various
components of $\vec{E}_1$ and their derivatives.  Eliminating
$E_{1\phi}^{'}$, $E_{1\phi}^{''}$, $E_{1z}^{'}$ and $E_{1z}^{''}$
using equations (\ref{Ephip}), (\ref{Ephipp}), (\ref{Ezp}) and the
derivative of (\ref{Ezp}), we obtain a differential equation for
$E_{1R}(R)$:
\begin{equation}
D_4E_{1R}^{'} + D_5E_{1z} + D_6E_{1R} = 0,
\label{ERp}
\end{equation}
where $D_4$, $D_5$ and $D_6$ are again functions of $R$ and are given
in Appendix~\ref{sec_coeffs}.

We have thus reduced the linear mode analysis problem to a pair of
first-order differential equations, (\ref{Ezp}), (\ref{ERp}), for
$E_{1z}(R)$ and $E_{1R}(R)$.  For convenience, we write down the two
equations again:
\begin{equation}
E_{1z}^{'} = -{D_2\over D_1}E_{1z} - {D_3\over D_1}E_{1R},
\label{difeq1}
\end{equation}
\begin{equation}
E_{1R}^{'} = - {D_5\over D_4}E_{1z} - {D_6\over D_4}E_{1R}.
\label{difeq2}
\end{equation}
These equations constitute an eigenvalue problem, where $\omega$ is
the eigenvalue.  By numerically solving the equations with appropriate
boundary conditions, we obtain $\omega$ for given values of $m$ and
$k$.

The singular points of the equations are located at the radii where
$D_1(R)$ and $D_4(R)$ vanish.  Anticipating later discussion, we write
down here the expression for the quantity $D_1 D_4$:
\begin{equation}
D_1D_4 =-{g\over\omega R}\left[{(\omega Rf-ckRh)^2+(\omega Rg-mch)^2-(ckRg-mcf)^2
\over (\omega Rf-ckRh)}\right].
\label{D1D4}
\end{equation}
Also, from equation (\ref{vdrift}), the perturbed velocity is
\begin{equation}
\frac{\vec{v}_1}{c} = \frac{\vec{E}_1\times\vec{B}_0
+\vec{E}_0\times\vec{B}_1}{B^2},
\end{equation}
which in component form gives
\begin{eqnarray}
\frac{v_{1R}}{c} &=& \frac{gE_{1\phi}+fE_{1z}}{f^2+g^2}, \label{v1R}\\
\frac{v_{1\phi}}{c} &=& -\frac{gE_{1R}-hB_{1z}}{f^2+g^2}, \label{v1phi}\\
\frac{v_{1z}}{c} &=& -\frac{fE_{1R}+hB_{1\phi}}{f^2+g^2}. \label{v1z}
\end{eqnarray}

We now consider boundary conditions.  A physically valid perturbation
will be well-behaved on the axis ($R=0$) and will satisfy suitable
boundary conditions at the jet boundary ($R=R_j$).  The condition on
the axis is different for axisymmetric ($m=0$) and non-axisymmetric
($|m|\geq1$) perturbations, so we consider each of these cases
in turn.  At the jet boundary, we assume that the jet is constrained
by a ``rigid wall'' and we write down the corresponding boundary
condition.  In the following, we employ the specific forms of $f(R)$,
$g(R)$, $h(R)$ given in equations (\ref{B0phi}), (\ref{B0z}),
(\ref{E0R}).

\subsection{Boundary Condition on the Axis: $m=0$}

Setting $m=0$ and substituting the expressions for $f(R)$, $g(R)$,
$h(R)$ in $D_1-D_6$, we find that the leading terms of the
differential equations (\ref{difeq1}), (\ref{difeq2}) at small $R$
are given by
\begin{eqnarray}
E_{1z}^{'} &=& {a_{zz} \over R} E_{1z} + a_{zR} E_{1R}, \\
E_{1R}^{'} &=& {a_{Rz} \over R^2} E_{1z} + {a_{RR}\over R} E_{1R},
\end{eqnarray}
where
\begin{eqnarray}
a_{zz} &=& -{2\omega\over ck}, \\
a_{zR} &=& {i(c^2k^2-\omega^2)\over c^2k}, \\
a_{Rz} &=& {4i\omega\over k(ck-\omega)}, \\
a_{RR} &=& {(ck+2\omega)\over ck}.
\end{eqnarray}

Requiring the perturbation to be analytic as $R\to0$ immediately
gives the following solution near the axis,
\begin{eqnarray}
E_{1z} &=& K R^2, \label{E1zm0}\\
E_{1R} &=& -K {2ic\over (ck-\omega)} R,
\end{eqnarray}
where $K$ is an arbitrary normalization constant.

\subsection{Boundary Condition on the Axis: $|m|>0$}

When $m\neq 0$, we obtain
\begin{eqnarray}
a_{zz} &=& 0, \\
a_{zR} &=& {imA(ck-\omega)\over (Acm-\omega R_m)}, \\
a_{Rz} &=& -{im(Acm-\omega R_m) \over A(ck-\omega)}, \\
a_{RR} &=& -1,
\end{eqnarray}
where the constant $A$ is defined to be
\begin{equation}
A = \left[2(\gamma_m^2-1)\right]^{1/2}.
\label{Adef}
\end{equation}
The physically relevant solution close to the axis is then
\begin{eqnarray}
E_{1z} &=& K R^{|m|}, \label{E1zm}\\
E_{1R} &=& -K\, {\rm sgn}(m)\, {i(Acm-\omega R_m) \over 
A(ck-\omega)}\, R^{|m|-1}.
\end{eqnarray}

\subsection{Boundary Condition at the Jet Boundary: Rigid Wall}

We assume that our cylindrical jet is terminated at $R=R_j$ by a rigid
impenetrable wall.  By impenetrable we mean that no energy flows
across this boundary, either out of or into the jet, i.e., the
Poynting flux lies in the $\phi-z$ plane.  Equivalently, the velocity
vector has no radial component.

The equilibrium Poynting flux of course lies in the $\phi-z$ plane.
The perturbed Poynting flux is proportional to
$\vec{E}_1\times\vec{B}_0 +\vec{E}_0\times\vec{B}_1$.  Since
$\vec{E}_0$ is parallel to $\hat{R}$, the term
$\vec{E}_0\times\vec{B}_1$ is automatically in the $\phi-z$ plane.
The term $\vec{E}_1\times\vec{B}_0$ will also be in this plane if
$\vec{E}_1$ is precisely radial, i.e., both $E_{1\phi}$ and $E_{1z}$
vanish.  By equation (\ref{Ephi}), $E_{1\phi}$ is proportional to
$E_{1z}$.  We thus obtain the following boundary condition at the
outer wall:
\begin{equation}
E_{1z} = 0, \qquad R=R_j.
\label{bcout}
\end{equation}

\section{Numerical Results}

We have computed frequencies of modes by numerically solving the
differential equations (\ref{difeq1}) and (\ref{difeq2}), along with
the boundary conditions described in \S\S3.2--3.4.  For each choice of
$k$ and $m$, a countable infinity of solutions exists which may be
ordered by the number of zeros of ${\rm Re}[E_{1z}(R)]$, not counting zeros
at the boundaries.\footnote{${\rm Re}()$ stands for the real part of a
complex quantity.} The lowest-order solution (the ``fundamental
mode'') is such that ${\rm Re}[E_{1z}(R)]$ has no zeros between $R=0$ and
$R=R_j$, the next solution has one zero inside the jet, and so on.  In
the following we identify each mode by its radial mode number $n$
which we define to be the number of zeros.  As one might expect, the
mode frequency increases with increasing $n$.

We solve for the frequencies via a shooting method.  We start with a
guess value of $\omega$, make use of the expressions given in \S3.2 or
\S3.3 (depending on the value of $m$) to set up the initial solution
at small $R$, and integrate equations (\ref{difeq1}) and
(\ref{difeq2}) to $R=R_j$.  We then adjust $\omega$ in the complex
plane until the outer boundary condition given in \S3.4 is satisfied.
The only subtle point is that the quantities $D_1$ and $D_4$ appear in
the denominators of various coefficients in equations (\ref{difeq1})
and (\ref{difeq2}), and so their zeros correspond to poles in the
solution.  To avoid these poles, we treat $R$ as a complex variable
and integrate the equations over a ``safe'' trajectory in complex-$R$
space.  Since the solution is analytic, the exact track that we follow
is unimportant so long as it lies {\it above} all singularities in the
$R$-plane.  \citet{ip96} give a detailed discussion of this topic in
connection with current-driven instabilities in force-free jets.  The
reader is also referred to standard discussions of this point in
plasma physics texts in the context of Landau damping, or
\citet*{ggn86} for a discussion in the context of accretion disk
instabilities.

\subsection{Axisymmetric Modes: $m=0$}

\begin{figure}
\includegraphics[width=\columnwidth]{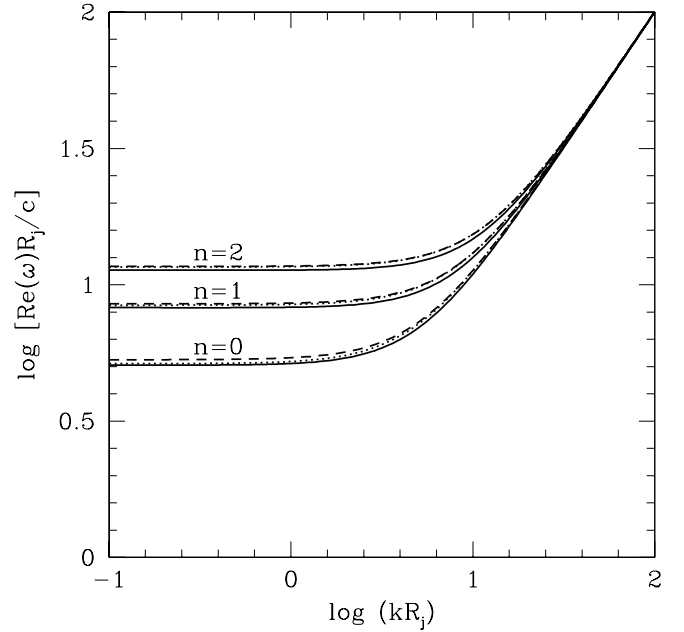}
\caption{Dispersion relation for axisymmetric modes ($m=0$) in Model A
(solid lines), Model B (dashed lines) and Model C (dotted lines).
From below the curves correspond to radial mode numbers $n=0$, 1, 2.
All the modes are stable.}
\label{fig3}
\end{figure}

Figure \ref{fig3} shows the dispersion relation --- the variation of $\omega$
with $k$ --- of axisymmetric modes ($m=0$).  Results are shown for
Models A, B and C (eqs. \ref{ModelA}, \ref{ModelB}, \ref{ModelC}) for
three radial mode numbers: $n=0$, 1, 2.  For all $k$, we find that the
mode frequency is real, which means that all these modes are stable.

For large values of $kR_j$, the mode frequency asymptotes to
$\omega=\pm ck$, so the modes behave like electromagnetic waves moving
parallel or anti-parallel to the $z$-axis.  At small $k$, however, the
frequency asymptotes to a constant value.  There is thus a minimum
frequency for propagating modes inside the jet.  The minimum frequency
is of order the inverse of the light-crossing time across a radial
wavelength of the mode (e.g., $\omega_{\rm min} \sim 2\pi c/R_j$ for
the mode with $n=0$).

We find that the dispersion relations of modes with positive and
negative $k$ are not quite the same.  The difference arises because
the background has a non-zero velocity in the $z$ direction, which
breaks the symmetry between waves propagating towards $+z$ and $-z$.
The effect is, however, quite weak.

\subsection{Non-Axisymmetric Modes: $m=\pm1$}

The most interesting modes are those with $m=\pm1$.  These modes are
stable in Model A, but unstable in Models B and C.

\begin{figure}
\includegraphics[width=\columnwidth]{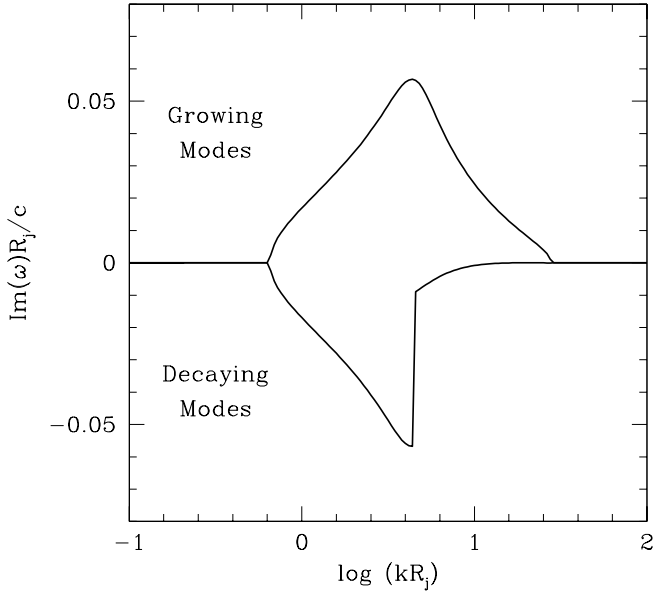}
%\vskip -0.5in
\caption{Imaginary part of $\omega$ for modes in Model B with $m=1$,
$n=0$.  Growing modes have ${\rm Im}(\omega)>0$, while decaying modes have
${\rm Im}(\omega)<0$.}
\label{fig4}
\end{figure}

Figure \ref{fig4} shows ${\rm Im}(\omega)$\footnote{${\rm Im}()$
refers to the imaginary part of a complex quantity.} as a function of
$kR_j$ for a sequence of modes in Model B; the modes correspond to
$m=1$, $n=0$.  In this sequence, modes with $kR_j < 0.65$ and those
with $kR_j>28$ are stable and have $\omega$ real.  However, for $0.65
< kR_j < 28$, we find a pair of modes with complex values of $\omega$.
The branch with ${\rm Im}(\omega)>0$ corresponds to growing modes, and
the branch with ${\rm Im}(\omega)<0$ to decaying modes.\footnote{We
discuss the sudden jump in the value of ${\rm Im}(\omega)$ for the
decaying branch at the end of \S4.2.}

\begin{figure}
\includegraphics[width=\columnwidth]{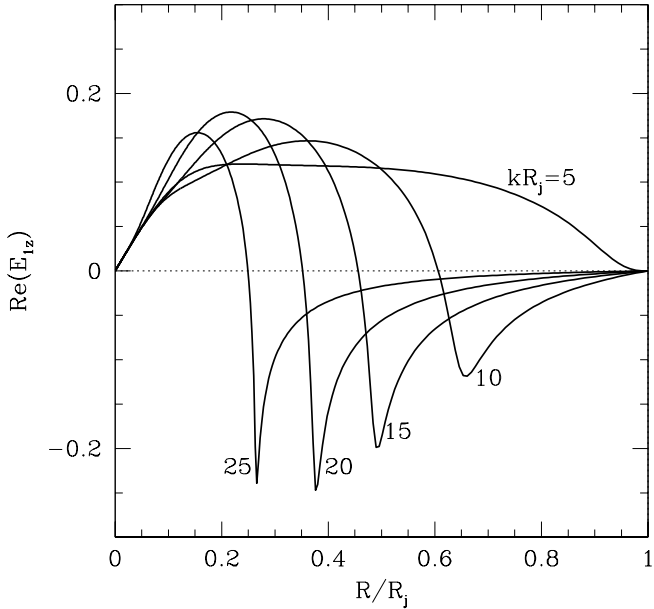}
%\vskip -0.5in
\caption{Eigenfunctions corresponding to growing modes in Model B with $m=1$,
$n=0$ and $kR_j=5$, 10, 15, 20 and 25.  The real part of $E_{1z}$ is
plotted.}
\label{fig5}
\end{figure}

Figure \ref{fig5} shows eigenfunctions corresponding to a few of the
growing modes.  Plotted is ${\rm Re}(E_{1z})$ as a function of the scaled
radius $R/R_j$.  The mode corresponding to $kR_j=5$ is representative
of all modes with $kR_j\ \lesssim\ 5$.  These modes have
eigenfunctions with no zero crossings between $R=0$ and $R=R_j$.  By
our definition, the modes correspond to $n=0$.  Each of the remaining
eigenmodes in Fig. \ref{fig5} has a pronounced dip in ${\rm Re}(E_{1z})$
which causes a zero crossing.  These dips result from a singularity in
the equations, as we discuss below.  If we discount the
singularity-induced zero crossings, then these eigenfunctions may also
be identified as $n=0$ modes.

\begin{figure}
\includegraphics[width=\columnwidth]{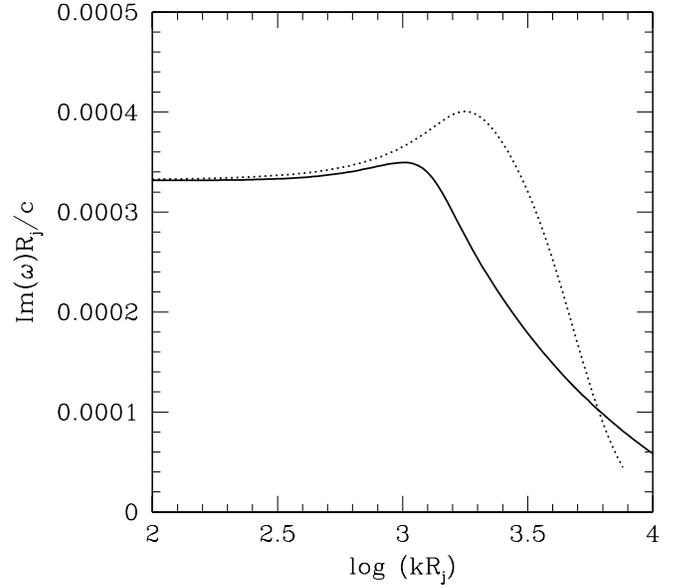}
%\vskip -0.5in
\caption{Imaginary part of $\omega$ for growing modes in Model C.  The
solid line corresponds to modes with $m=1$, $n=0$ and the dotted line
corresponds to modes with $m=-1$, $n=0$.}
\label{fig6}
\end{figure}

\begin{figure}
\includegraphics[width=\columnwidth]{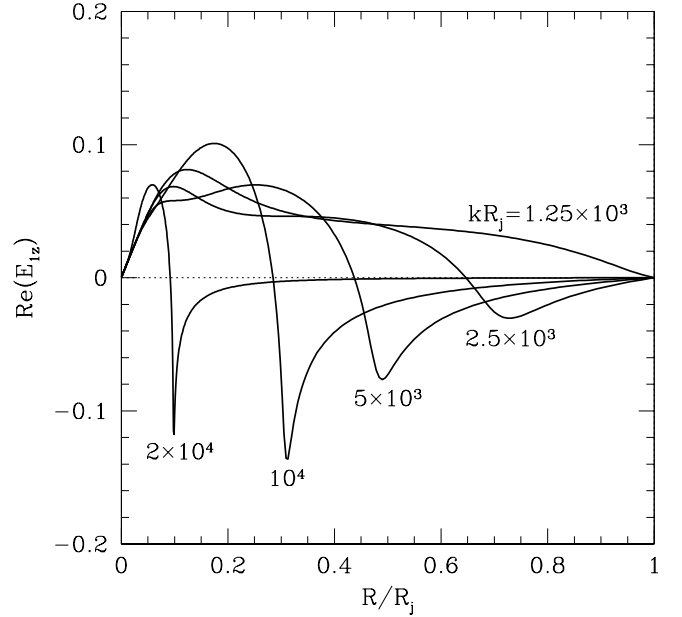}
%\vskip -0.5in
\caption{Eigenfunctions corresponding to growing modes in Model C with $m=1$,
$n=0$ and $kR_j=1.25\times10^3$, $2.5\times10^3$, $5\times10^3$,
$10^4$ and $2\times10^4$.  The real part of $E_{1z}$ is plotted.}
\label{fig7}
\end{figure}

Figures \ref{fig6} and \ref{fig7} show similar results for Model C.
Growing modes (and their decaying counterparts) are present for $m=1$
and all $kR_j < 2.1\times 10^4$.  Modes with $m=-1$ are also unstable
(see Fig. \ref{fig6})\footnote{In the case of Model B, modes with
$m=-1$ appear to be stable, and only the $m=+1$ modes show an
instability.}.  Figure \ref{fig7} shows a few eigenfunctions.  All
modes with $kR_j \lesssim 1.3\times10^3$ have eigenfunctions with the
standard $n=0$ shape (see the mode with $kR_j=1.25\times10^3$ in
Fig. \ref{fig7}).  For larger values of $k$, the eigenfunctions
develop negative spikes due to the presence of a singularity (see
Fig. \ref{fig7}).  However, we still view them as $n=0$ modes.

We have determined numerically that the singularities which cause the
dips in the eigenfunctions are due to zeros in the function $D_4(R)$
defined in \S3.1.  This function appears in the denominator of the
differential equation (\ref{difeq2}), and hence its zeros behave like
poles.\footnote{In contrast, although the function $D_1(R)$ appears in
the denominator of equation (\ref{difeq1}), its zeros do not cause a
real singularity since the terms $D_2$ and $D_3$ also go to zero at
the same locations.}

\begin{figure}
\includegraphics[width=\columnwidth]{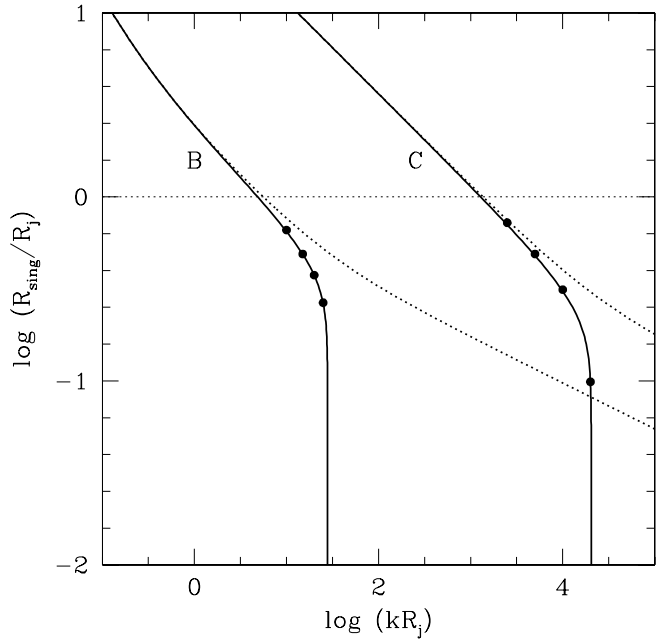}
%\vskip -0.5in
\caption{Location of the singularity $\rsing$ as a function of $kR_j$
in Models B and C, calculated using eq. (\ref{Rsing}).  Solid curves
correspond to modes with $m=1$, $n=0$ and dotted lines to modes with
$m=-1$, $n=0$.  Solid dots show the locations of minima in the
eigenfunctions plotted in Figs. \ref{fig5} and \ref{fig7}.}
\label{fig8}
\end{figure}

Equation (\ref{D1D4}) gives the analytic form of the quantity
$D_1D_4$.  Since the modes of interest to us have ${\rm Re}(\omega)$ very
nearly equal to $ck$, let us substitute $\omega=ck$ in this equation.
Then, setting $D_1D_4$ equal to zero gives the following relation
between the wavenumber $k$ of the mode and the radius $\rsing$ of the
singularity:
\begin{equation}
k\rsing = \frac{m[f(\rsing)+h(\rsing)]}{g(\rsing)+{\rm sgn}(m)
\sqrt{g^2(\rsing) +[f^2(\rsing)-h^2(\rsing)]}}.
\label{Rsing}
\end{equation}
Figure \ref{fig8} shows the position of the singularity $\rsing$ as a
function of $kR_j$ for modes with $m=\pm1$ in Models B and C, as
calculated with this equation.  For comparison, the dots show the
radii at which the functions ${\rm Re}[E_{1z}(R)]$ reach their minima in
the eigenfunctions plotted in Figs. \ref{fig5} and \ref{fig7}.  The
agreement between the analytical curve and the dots is excellent,
showing that equation (\ref{Rsing}) captures the physics of the
singularity.

From Fig. \ref{fig8} we see that the singularity lies inside the jet
($\rsing<R_j$) only for a finite range of $k$ above a certain minimum
value.  For values of $k$ smaller than this minimum, the singularity
is outside the jet (for very small $k$ it is well outside the jet).
In the case of Model B, the singularity enters the jet from outside
when $kR_j\sim5$ and it disappears (for $m=+1$) at $R=0$ when
$kR_j\sim28$.  This is the primary range of $k$ over which an unstable
mode is present.  At $kR_j\sim28$, the singularity is barely present
near the center of the jet and we have a very weakly growing mode.
With decreasing $k$, the singularity moves outward and the growth rate
of the mode increases (Fig. \ref{fig4}).  At $kR_j\sim5$, when the
singularity reaches the wall, the growth rate is close to its maximum
value.  At yet smaller values of $k$, the singularity moves outside
the outer wall, but its presence is still felt and there is continued
instability.  The growth rate however decreases with decreasing $k$.

A similar pattern is seen in Model C.  Unstable modes are present only
for $kR_j \lesssim2\times10^4$.  With decreasing $k$ the growth rate
increases and reaches its maximum approximately when the singularity
reaches the jet boundary ($\rsing=R_j$), which happens at
$kR_j\sim1.3\times10^3$.  In contrast to Model B, however, the growth
rate remains large even for smaller values of $k$, and the instability
survives down to $k\to0$.

We finally discuss the peculiar behavior of ${\rm Im}(\omega)$ in the
decaying branch of modes in Fig. \ref{fig4}.  As we mentioned earlier, in
numerically solving for the eigenvalue we must integrate the
differential equations (\ref{difeq1}) and (\ref{difeq2}) along a path
in the complex-$R$ plane that lies above the poles in the solution.
For growing modes, the pole is located below the real $R$-axis.  We
can therefore integrate along the real $R$-axis without any
difficulty.  For decaying modes, the pole is above the real $R$-axis
and now we must choose the integration path with care.  If the
singularity has ${\rm Re}(\rsing)>R_j$, i.e., the singularity is outside
the jet, there is no problem and we can simply integrate along the
real axis.  However, when $0<{\rm Re}(\rsing)<R_j$, we have to deform the
integration path.  In our calculations, we integrate from $R=0$ along
a path with ${\rm Im}(R)={\rm Re}(R)$ until the point ${\rm Im}(R)={\rm Re}(R)=R_j$ and we
then integrate down to $R=R_j$.  The jump in ${\rm Im}(\omega)$ in Fig. \ref{fig4}
is the result of the singularity moving into the jet.  To the left of
the break, the singularity is located at $R>R_j$.  Here the
eigenvalues of the growing and decaying modes are complex conjugates of
each other.  However, to the right of the break, the singularity has
moved inside the jet ($R<R_j$) and now the complex conjugate symmetry
is broken.

We note that eigenfunctions and eigenvalues of decaying modes are not
very meaningful.  This can be shown from an initial condition analysis
along the lines of Landau's treatment of plasma damping.  The reader
is referred to \citet{ip96} for a detailed discussion of this
topic.

\subsection{Why $m=\pm1$ is Special}

We have not exhaustively explored modes with $|m|>1$.  However, in
spot tests with various choices of $m$, $n$ and $kR_j$ in Models A, B
and C, all modes were found to be stable.  We believe that, if at all,
there are only weakly unstable modes for $|m|>1$; there is no sign of
the kind of vigorous instability described in \S4.2 for modes with
$m=\pm1$.  So why is $m=\pm1$ special?  The answer to this question is
well-known in the magnetic confinement literature (e.g.,
\citealt{bat78}).  We discuss it briefly here for completeness.

Consider the radial component of the perturbed velocity $v_{1R}$ near
the axis of the jet.  Equation (\ref{v1R}) gives the expression for
$v_{1R}$ in terms of the perturbed electric field components $E_{1z}$
and $E_{1\phi}$, and equation (\ref{Ephi}) shows the relation between
these two field components.  For small values of $R$ near the axis, we
have
\begin{equation}
g(R) \approx 1, \quad
f(R) \approx h(R) = A\frac{R}{R_m},
\end{equation}
where the quantity $A$ is defined in equation (\ref{Adef}), and
\begin{equation}
C_1 \approx \frac{(\omega R_m-Acm)}{AR(\omega-ck)}.
\end{equation}

Consider first modes with $m=0$.  Equation (\ref{E1zm0}) shows that
$E_{1z}\approx KR^2$ near the axis.  Substituting this in equation
(\ref{v1R}) and using the other approximations given above, we find
\begin{equation}
v_{1R} \approx \frac{\omega R_m}{A(\omega-ck)}KR+{\cal O}(R^3).
\end{equation}
By symmetry, the velocity goes to zero on the axis, and the flow
consists of a simple radial divergence.

Consider next modes with $m\neq0$.  Using equation (\ref{E1zm}) for
$E_{1z}$, we find
\begin{equation}
v_{1R} \approx \frac{(\omega R_m-Acm)}{A(\omega-ck)}KR^{|m|-1}
\cos m\phi + {\cal
O}(R^{|m|+1}),
\label{v1R0}
\end{equation}
where we have included $\cos m\phi$ to show the angular dependence of
the mode.  The leading term goes like $R^{|m|-1}$, which corresponds
to $R^0$ when $m=\pm1$.  This means that the mode has a finite radial
velocity, and hence a finite radial displacement, on the axis when
$|m|=1$.  The $\cos\phi$ dependence of $v_{1R}$, coupled with the fact
that $v_{1\phi}$ has the same amplitude but a $\sin\phi$ dependence,
ensures that the velocity vector is unique and analytic at $R=0$.  By
writing the velocity vector in cartesian coordinates, it is easily
seen that the complex phase of $v_{1R}$ determines the orientation of
the velocity vector in the $xy$-plane.  If we consider values of
$|m|\geq2$, the velocity vanishes on the axis, just as in the case of
$m=0$.

This then reveals what is special about $|m|=1$ modes.  These are the
only modes in which fluid perturbations communicate across the axis
and cause the jet to shift bodily across the axis.  In modes with
$|m|=1$ the center of mass of the jet itself shifts into a spiral
shape, which is the characteristic feature of the kink or screw mode.
For all other values of $|m|$, the center of mass remains on the axis
and the perturbations are concentrated on the outside.

In helical MHD configurations in the laboratory, the $|m|=1$ kink mode
is known to be highly unstable and to be the greatest threat to the
stability of equilibria \citep{bat78}.  Not surprisingly we see the
same feature in our force-free jet equilibria.

\subsection{Growth Rate of the Instability}

The growth rate of the fastest growing mode is a matter of practical
interest since it limits the lifetime of an unstable system.  As
discussed in \S4.2, for the models we have considered here, the most
unstable mode generally has a singularity close to the outer wall:
$\rsing\sim R_j$.  Knowing this, we estimate here the fastest growth
rate by assuming that the pole is located at $\rsing=1.1R_j$.  (We
locate the singularity slightly outside the jet, since this speeds up
the numerical integrations considerably.)

Given an assumed value of $\rsing$, we can substitute this value in
equation (\ref{Rsing}) and make use of the expressions for $f(R)$,
$g(R)$, $h(R)$ given in \S2.2.  Recalling that Models B and C are in
the regime described in \S2.2.2, we note that $f^2(R_j)-h^2(R_j) \gg
g^2(R_j)$.  In addition, $f$ and $h$ are nearly equal to each other
and $\gamma_j$ is given by equation (\ref{gammaj2}).  We then find
that $kR_j\approx1.6\gamma_j$.  Also, the real part of the frequency
is nearly equal to $ck$.  Thus, we estimate
\begin{equation}
{\rm Mode~with~\rsing=1.1R_j}:\quad k\approx\frac{1.6\gamma_j}{R_j},
\quad {\rm Re}(\omega)\approx \frac{1.6\gamma_jc}{R_j}.
\label{modek}
\end{equation}
These estimates should apply to the fastest-growing mode.  The mode
with the maximum growth rate in Model B has ${\rm Re}(\omega)\approx
kR_j=4.4$.  Since Model B has $\gamma_j=2.5$, equation (\ref{modek})
predicts $kR_j\approx4.0$, which is close.  Similarly, the mode with
the maximum growth rate in Model C has ${\rm Re}(\omega)\approx
kR_j=1000$, whereas equation (\ref{maxgrowth}) with $\gamma_j=660$
predicts $kR_j\approx1060$.  We see that the approximate formula
(\ref{modek}) is quite good.

\begin{figure}
\includegraphics[width=\columnwidth]{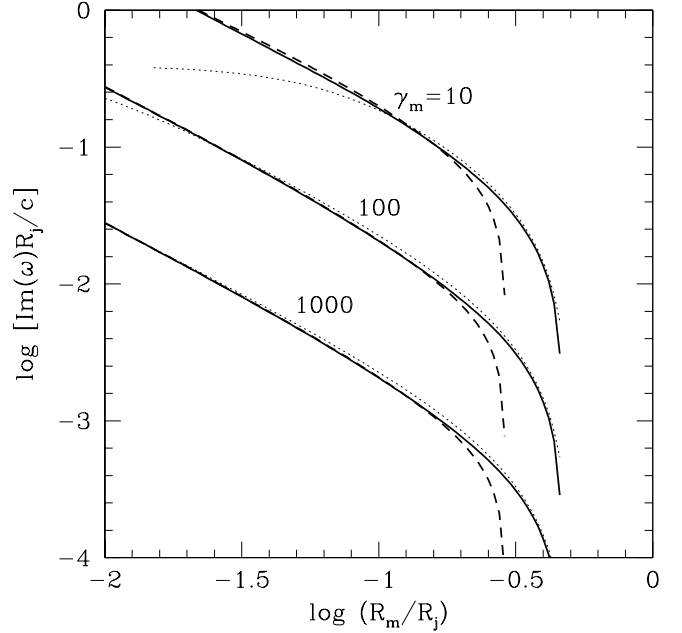}
%\vskip -0.5in
\caption{Numerically calculated growth rates of modes with $m=1$ and
$\rsing=1.1R_j$.  These modes have among the largest growth rates.
The solid lines show the results for a series of models for fixed
$\gamma_m$ and varying $R_m/R_j$.  The dotted lines are the growth
rates predicted by eq. (\ref{maxgrowth}).  Note the very good
agreement except near the top of the plot, where the models are
non-relativistic.  The dashed lines are the numerical growth rates for
modes with $k=0$.}
\label{fig9}
\end{figure}

\begin{figure}
\includegraphics[width=\columnwidth]{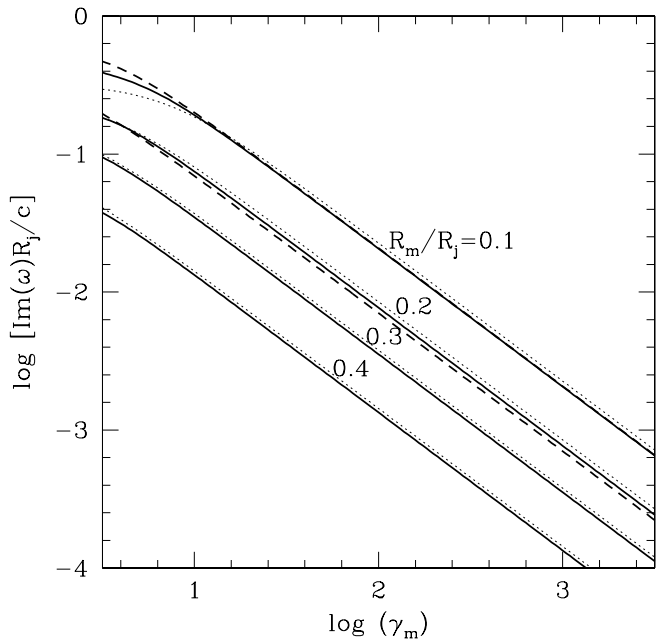}
%\vskip -0.5in
\caption{Numerically calculated growth rates of modes with $m=1$ and
$\rsing=1.1R_j$.  The solid lines show the results for a series of
models with fixed $R_m/R_j$ and varying $\gamma_m$.  The dotted lines
are the growth rates predicted by eq. (\ref{maxgrowth}).  The dashed
lines are the numerical growth rates for modes with $k=0$.}
\label{fig10}
\end{figure}

Our numerical results indicate that the growth rate ${\rm Im}(\omega)$ of
the fastest growing mode is proportional to ${\rm Re}(\omega)/\gamma_j^2$.
We also know that unstable modes are present only when $R_m < R_j$;
for instance, Model A with $R_m=1.2R_j$ has no unstable modes, whereas
Model B with $R_m=0.3R_j$ and Model C with $R_m=0.18R_j$ both have
unstable modes.  With these clues in mind, we obtain the following
empirical estimate for the growth rate of the fastest-growing mode:
\begin{equation}
{\rm Mode~with~\rsing=1.1R_j}:\quad {\rm Im}(\omega)\approx \frac{0.4}{\gamma_j}
\left(1-\frac{2R_m}{R_j}\right)\frac{c}{R_j}.
\label{maxgrowth}
\end{equation}
The coefficients 0.4 and 2 are very approximate (to emphasize this, we
give only the leading digit for each).  Nevertheless, as we show in
Figs. \ref{fig9} and \ref{fig10}, this approximate formula does quite
a good job of fitting the numerical results for a wide range of
models.  The only region of parameter space where the formula fails is
when the underlying equilibrium becomes ``non-relativistic'' and
$\gamma_j$ approaches unity.  These models are near the upper end of
Figs. \ref{fig9} and \ref{fig10} and have extremely large growth
rates.

Although modes with $\rsing\sim R_j$ have the largest growth rates,
these modes have relatively short wavelengths $\ll R_j$ along the
$z$-axis (see eq. \ref{modek}).  With such short wavelengths it is not
clear if these instabilities can grow to large amplitude.  It is
therefore interesting to consider modes with $k\to0$.  Figure
\ref{fig4} shows that Model B is stable as $k\to0$, whereas
Fig. \ref{fig6} indicates that the $k=0$ mode in Model C is nearly as
unstable as the fastest-growing mode.

The dashed lines in Figs. \ref{fig9} and \ref{fig10} show numerical
results for the growth rates of modes with $k=0$ for various
combinations of the model parameters $\gamma_m$ and $R_m$.  For small
values of $R_m\lesssim 0.1R_j$, the results are nearly identical to
those we described above for the fastest-growing mode
($\rsing=1.1R_j$, solid lines).  This is to be expected based on the
results shown in Fig. \ref{fig6} for Model C, which has $R_m=0.1R_j$.
With increasing $R_m$, however, the $k=0$ modes become less unstable
than the modes with $\rsing\sim R_j$.  By $R_m\sim0.3R_j$, the $k=0$
modes are fully stable, thus explaining the result shown in Model B
(Fig. \ref{fig4}), which has $R_m=0.3R_j$.

\subsection{Spatial Growth of Unstable Modes}

The discussion so far was limited to modes with real $k$ and complex
$\omega$.  An equally interesting problem is to consider modes with
real $\omega$ and complex $k$.  From equation (\ref{eigenfunction}),
we see that the eigenfunctions take the form
\begin{eqnarray}
\vec{E}_1 &=& [E_{1R}(R)\hat{R}+E_{1\phi}(R)\hat{\phi}+E_{1z}(R)\hat{z}] \nonumber \\
    &\times&      \exp[-i\omega t + im\phi+i{\rm Re}(k)z-{\rm Im}(k)z] \nonumber \\
    &\propto& \exp[i{\rm Re}(k)z]\exp(z/Z),
\end{eqnarray}
where $\omega$ is real and $Z=-1/{\rm Im}(k)$ is the scale length on which
the mode $e$-folds in the $z$-direction.  Such spatially growing
``convected'' modes are particularly relevant for sources with
long-lived steady jets.

As discussed in \citet{pc85} and \citet{ac92},
there is a strong symmetry between modes with real $k$ and complex
$\omega$, and those with complex $k$ and real $\omega$.  In
particular, the growth rates of the two kinds of modes are related by
\begin{equation}
{\rm Im}(k) = -{\rm Im}(\omega)/v_g,
\end{equation}
where $v_g=\partial{\rm Re}(\omega)/\partial{\rm Re}(k)$ is the group velocity
of the mode.  Our unstable $m=1$ modes have $v_g$ very nearly equal to
$c$.  Therefore, we immediately obtain from equation
(\ref{maxgrowth}) the following estimate of the spatial $e$-folding
scale $Z$ of the fastest-growing convected mode:
\begin{equation}
{\rm Mode~with~\rsing=1.1R_j}:\quad \frac{Z}{R_j}\approx
\frac{2.5 \gamma_j}{(1-2R_m/R_j)}.
\label{maxgrowth2}
\end{equation}
Zero-frequency modes ($\omega=0$) should have almost the same $Z$ for
small values of $R_m/R_j$, but the growth should cut off at a somewhat
smaller value of $R_m$ compared to the modes with $\rsing=1.1R_j$.

\begin{figure}
\includegraphics[width=\columnwidth]{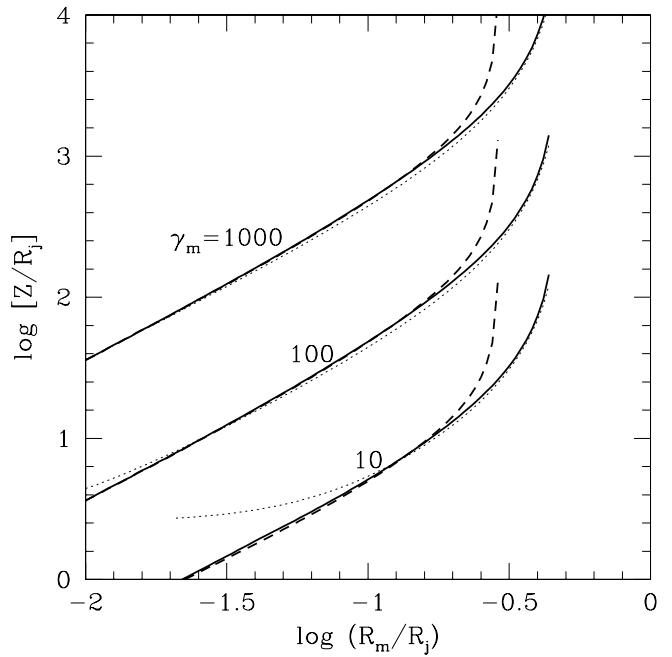}
%\vskip -0.5in
\caption{Numerically calculated $e$-folding scale $Z$ for modes with
$m=1$, $\omega$ real and $\rsing=1.1R_j$.  These modes have among the
largest largest growth rates.  The solid lines show the results for a
series of models with a given value of $\gamma_m$ and different values
of $R_m/R_j$.  The dotted lines are the growths predicted by
eq. (\ref{maxgrowth2}).  Note the very good agreement except near the
bottom of the plot, where the models are non-relativistic.  The dashed
lines are the numerical values of $Z$ for modes with $\omega=0$.}
\label{fig11}
\end{figure}

Figure \ref{fig11} shows numerical results.  Modes with $\rsing=1.1R_j$ have
growths consistent with equation (\ref{maxgrowth2}), and the modes
with $\omega=0$ have similar growths except that the instability cuts
off at somewhat smaller values of $R_m$.  The results are as expected
and are very similar to those shown in Fig. \ref{fig9}.

The above results correspond to an idealized cylindrical jet.  In the
case of real jets we must allow for a finite opening angle $\theta_j
\equiv dR_j/dz$.  Many force-free jet models have opening angles that
vary inversely as the Lorentz factor: $\theta_j \sim {\rm
few}/\gamma_j$ (\citetalias{tchekhovskoy_ff_jets_2008}).  Using this scaling we can estimate
approximately the evolution of the mode amplitude $a$ with distance:
\begin{eqnarray}
\frac{da}{dR_j}=\frac{1}{\theta_j}\frac{da}{dz}
  &\approx& \frac{\gamma_j}{\rm few}\,\frac{da}{dz} \approx
  \frac{\gamma_j}{\rm few}\,\frac{a}{Z} \nonumber\\ 
  &\approx& 
  \frac{\gamma_j}{\rm few}\,\frac{a}{2.5\gamma_jR_j} \approx
  \frac{1}{\rm few\times2.5}\,\frac{a}{R_j}.
\end{eqnarray}
Solving this differential equation, and using $R_j\propto
z^{\alpha/4}\sim z^{0.5-1}$ (eq. \ref{Rjz}), we obtain
\begin{equation}
a(z) \propto z^\epsilon,
\label{avsz}
\end{equation}
where $\epsilon$ is a small number $\lesssim0.1$.  This estimate is
very crude, but it does suggest that, in realistic jets, the unstable
kink mode we have studied in this paper grows only weakly with
increasing distance.

\subsection{Towards an Improved Instability Criterion}

In \S1, we introduced three different instability criteria, of which
the IPL and TMT criteria refer specifically to rotating force-free
jets.  Since $B_z$ is practically constant in our equilibria, all of
our models are close to the boundary between stability and instability
according to the IPL criterion (eq. \ref{IPLcriterion}).  Similarly,
since $B_\phi^2 \approx \Omega^2R^2B_p^2$ in our equilibria, our
models are marginally stable according to the TMT criterion
(eq. \ref{TMTcriterion}).\footnote{A strict application of the TMT
criterion would indicate that our models are unstable, since $B_\phi^2
> E^2=\Omega^2R^2B_p^2$.  However, $B_\phi^2-E^2 \ll B_\phi^2$, so the
models deviate only slightly from marginal stability.}  It is thus not
possible to understand from either of these criteria why Model A is
stable and Models B and C are unstable.

It is, of course, not surprising that the above instability criteria
fail.  Our jet equilibria include the effects of poloidal field
curvature, which were not considered by the previous authors.  For
easier comparison with previous work, let us rewrite our balance
condition (\ref{balance2}) as follows:
\begin{equation}
{1\over R^2}{d\over dR}\left[{(B_{0\phi}^2-E_{0R}^2)R^2\over
8\pi}\right] = -{d\over dR}\left({B_{0z}^2\over
8\pi}\right)+\left({E_{0R}^2-B_{0z}^2\over 4\pi R_c}\right).
\label{balance3}
\end{equation}
If we leave out the last term, this is equivalent to equation (6) in
\citet{ip96} and equation (16) in \citet{lyu99}.  The IPL instability
criterion states that the quantity $dB_{0z}^2/dR$ should be negative.
We might wish to generalize this by saying that the right hand side of
equation (\ref{balance3}), including the poloidal curvature term,
should be positive.  Unfortunately, this simple modification is not
sufficient since the right hand side is positive for all of our
models, whereas not all our models are unstable; not only should the
right hand side be positive, its magnitude should be larger than some
amount.  The same seems to be true with the TMT criterion.  This
criterion indicates that all our equlibria should be unstable, whereas
only some of them are.

Qualitatively, what distinguishes the unstable Models B and C from the
stable Model A is that the former have made the transition to the
second accleration regime.  This is reflected in their $\gamma(R)$
profiles (Fig. 2) which have $d\gamma/dR<0$ at larger radii.  Thus,
one might guess that instability requires the jet to be in the second
acceleration regime and/or the jet to have a declining $\gamma(R)$.
Once again, these conditions by themselves are not sufficient.  To
have an instability, $\gamma(R)$ should decline over a sufficiently
broad range of radius, e.g., $R_m$ should be less than $\sim0.45R_j$
in our models.

An alternate approach which we have found useful is to focus on the
left-hand side of equation (\ref{balance3}).  From equations
(\ref{B0phi}), (\ref{B0z}), (\ref{E0R}), we see that for our
equilibria we have $B_{0\phi}^2-E_{0R}^2 =(R/R_m)^4 B_{0z}^2$.
Furthermore, we have seen that modes with $\rsing\sim R_j$ (the
fastest-growing modes) are unstable so long as $R_m \lesssim0.45 R_j$,
while modes with $k\to0$ (long-wavelength modes) are unstable for $R_m
\lesssim 0.3 R_j$.  From this, we obtain the following approximate
instability criteria:
\begin{eqnarray}
{\rm Modes~with~\rsing\sim R_j}:&~&	\left(B_{0\phi}^2-E_{0R}^2
\right)^{1/2} > 5 |B_{0z}|, \\
{\rm Modes~with}~k\to0:&~&	\left(B_{0\phi}^2-E_{0R}^2
\right)^{1/2} > 12 |B_{0z}|,
\end{eqnarray}
where we have set $R=R_j$ to obtain the numerical coefficients on the
right.  These conditions are easier to interpret if we boost to the
comoving frame of the fluid (V. Pariev, private communication), where
the electric field vanishes.  In this frame, the criterion for
instability becomes
\begin{eqnarray}
{\rm Modes~with~\rsing\sim R_j}:&~& |B_{0\phi,{\rm comov}}| > 5
|B_{0z,{\rm comov}}|, \label{comov1}\\ {\rm Modes~with}~k\to0:&~&
|B_{0\phi,{\rm comov}}| > 12 |B_{0z,{\rm comov}}|. \label{comov2}
\end{eqnarray}
That is, in the comoving frame, the toroidal field must dominate the
poloidal field by more than a certain critical factor.\footnote{Since
our equilibria assume a constant $\Omega$ for all field lines, the
criteria (\ref{comov1}) and (\ref{comov2}) are technically valid only
for such models.  However, since the criteria have been written
without any explicit reference to $\Omega$, they may be valid more
generally even when $\Omega$ varies with $R$.}  Written in this form,
the condition resembles the KS criterion (eq. \ref{KScriterion}).

\section{Summary and Discussion}

The relativistic jet model we have considered in this paper is
particularly simple: it is cylindrical, it assumes force-free
conditions, and it assumes rigid rotation.  Within the limitations of
these reasonable approximations, we have attempted to be as close to
numerically simulated jets as possible.  We include the effect of
poloidal field curvature, which is known to play an important role in
numerical force-free jets (\S2.1), and we choose functional forms for
the various field components in the equilibrium (\S2.2) to match as
closely as possible our previous force-free simulations (\citetalias{tchekhovskoy_ff_jets_2008}).

Our equilibrium model is described by two parameters: the maximum
Lorentz factor $\gamma_m$, and the radius at which this maximum is
achieved $R_m$.  The ratio of the latter to the jet radius $R_j$
determines the basic physics of the equilibrium.  Models in which
$R_m/R_j>1$ have $\gamma(R)$ increasing monotonically with radius $R$
out to some maximum Lorentz factor $\gamma_j<\gamma_m$ at the outer
edge of the jet.  Model A (Fig. \ref{fig2}) is an example.  In these
models the entire jet is in the first acceleration regime (see \S2.1,
2.2.1 and \citetalias{tchekhovskoy_ff_jets_2008} for details).  We find that all these models are
perfectly stable.

Models with $R_m/R_j<1$ are more interesting.  Here, $\gamma(R)$
increases upto a maximum value $\gamma_m$ at $R=R_m$ and then
decreases down to a Lorentz factor $\gamma_j<\gamma_m$ at $R=R_j$.
Models B and C (Fig. \ref{fig2}) are examples of this kind of model.
In these models, the jet fluid at $R<R_m$ is in the first acceleration
regime, while the fluid at $R_m<R<R_j$ is in the second acceleration
regime.  We find that the subset of these models with $R_m/R_j
\lesssim 0.45$ are linearly unstable.  For $R_m/R_j$ just below 0.45,
all the unstable modes have short wavelengths in the $z$-direction:
$\lambda =2\pi/k_z\sim 2\pi R_j/\gamma_j$.  With decreasing $R_m$, a
wider range of $k_z$ becomes unstable, and for $R_m/R_j \lesssim 0.3$,
we find that waves with $k_z= 0$, i.e., with arbitrarily long
wavelengths, are unstable.  The latter modes are perhaps of most
interest since they are likely to grow to the largest amplitudes.  The
numerical results are summarized in Figs. \ref{fig3}--\ref{fig11}.

The unstable modes we find are all kink modes with azimuthal
wavenumber $m=\pm1$.  These are non-axisymmetric modes in which the
jet is distorted helically.  A key feature is that, at each $z$, the
center of mass of the jet is shifted away from $R=0$.  It is
well-known that MHD configurations with toroidal fields are especially
susceptible to the kink mode \citep{bat78}, and our models follow this
trend.  However, because our equilibria both rotate and move
relativistically along $z$, the criterion for instability is different
from the usual KS criterion (eq. \ref{KScriterion}).

We find that the typical growth rate of the unstable kink mode in our
jet models is given by equation (\ref{maxgrowth}): the $e$-folding
time is of order $\gamma_j$ times the light-crossing time $R_j/c$
across the jet.  For convected modes with a real frequency, this
translates to an $e$-folding length scale of order $\gamma_j$ times
the jet radius $R_j$.  Since jets typically have opening angles
$\sim{\rm few}/\gamma_j$, the net result is that the unstable modes
grow only slowly with distance from the base of the jet
(eq. \ref{avsz}).  Of course, relativistic jets in astrophysical
sources propagate over many decades, so in principle even this slow
growth might lead to a large amplitude of the perturbation.
Nevertheless, the fact that the growth is very slow reduces the
seriousness of the kink instability.

Our jet equilibria turn out to be close to the boundary between
stability and instability according to either the IPL or TMT criterion
(eqs.~\ref{IPLcriterion}, \ref{TMTcriterion}), so these criteria are
not useful for interpreting the results.  In addition, since our
models include the effects of poloidal field curvature, they lie
outside the range of validity of the IPL and TMT criteria.  The most
useful instability criterion we have come up with is that, in the
comoving frame of the jet fluid, the tangential field should be an
order of magnitude or more larger than the poloidal field
(eqs.~\ref{comov1}, \ref{comov2}).  Expressed thus, the criterion is
similar to the KS criterion (\ref{KScriterion}), except that it should
be applied in the comoving frame and $z$ should be taken to be $\sim
R_j$.

All the work described here assumes a rigid wall enclosing the jet at
the boundary $R=R_j$.  We have done some calculations with a constant
pressure boundary and we find unstable modes with much larger growth
rates compared to the rigid wall case.  However, since we are dealing
with a force-free jet, it is not clear that a constant pressure
boundary is particularly meaningful.  For instance, if the pressure is
from a non-relativistic gaseous envelope or cocoon, the gas would have
substantial inertia and (we suspect) would probably behave to first
approximation like a rigid wall.

Various authors have discussed mechanisms by which instabilities might
be suppressed in astrophysical jets.  \citet[and references
therein]{hmn07} have shown that an external wind or cocoon can
stabilize the Kelvin-Helmholtz mode in MHD jets, though it is not
clear if this is relevant for force-free jets.  \citet*{mso08} show
that lateral expansion causes instabilities to grow more slowly.  In a
sense, we have already included this effect when we derived the growth
rate estimate given in equation (\ref{avsz}).  In addition, we note
that some of the growth suppression seen by \citeauthor{mso08} is
probably because expansion causes different parts of the jet to lose
causal contact with one other.  This is not an issue for force-free
models, where signals propagate at the speed of light.

It would be interesting to simulate numerically the unstable modes
described in this paper.  Apart from verifying the linear theory, such
calculations will reveal the non-linear development of the mode.  Does
the kink mode saturate at a finite amplitude and lead to a
more-or-less coherent helical pattern or does it destroy the initial
equilibrium?  This important question can be answered only with 3D
simulations.  Since the kink mode involves lateral motion of the jet
across the axis $R=0$, the numerical technique used must be flexible
enough to allow such motions (e.g., as described by \citealt{mb08}).

We conclude by reminding the reader that the work described here
refers to a particularly simple model of relativistic jets which is
based on the force-free approximation.  In real jets, once the flow
crosses the fast magnetosonic point, the inertia of the gas starts to
play a role and the force-free approximation is no longer valid (e.g.,
\citealt{tchekhovskoy_monopole09}).  In this regime, we must consider
the full MHD equations.

\ack

The authors thank Alison Farmer for assistance during the early stages
of this work and Jonathan McKinney for numerous helpful discussions
and comments on the paper.  This work was supported in part by NASA
grant NNX08AH32G.

\appendix

\section{Constancy of Poloidal Magnetic Field across Force-free Jets}
\label{sec_Bp_const}

Figure \ref{fig1} shows that in numerical simulations of force-free
jets $B_p$ hardly changes with~$R$.  In this Appendix we show that
this is a common feature of all jet solutions that smoothly connect to
a spinning compact object at the base. 

Consider the force balance equation~(\ref{balance}).  Sufficiently
near the compact object, where the jet is in the first acceleration
regime (see \S\ref{eq_structure_jets}), we can drop the terms
proportional to $R_c^{-1}$ and $(B_\phi^2-E^2)$ since in the first
acceleration regime $\gamma_1^2 \ll \gamma_2^2$ leading to $E^2/B_p^2
\ll R_c/R$ and $B_\phi^2-E^2\ll B_p^2$ (\citetalias{tchekhovskoy_ff_jets_2008}).  Then the force
balance equation~(\ref{balance}) simplifies to
\begin{equation}
\frac{d(B_p^2)}{dR} \approx 0.
\end{equation}     
Therefore, sufficiently near the compact object the poloidal field is
nearly constant,
\begin{equation}
B_p(R) \approx  {\rm const}.
\label{eq_Bp_const}
\end{equation}
Each field line is labeled by the amount of poloidal magnetic flux
$\Phi$ that it encloses.  Due to~(\ref{eq_Bp_const}) this flux can be
written simply as
\begin{equation}
\Phi \approx \pi B_p R^2.
\label{eq_Phi_BpRsq}
\end{equation}
These relations are valid throughout the first acceleration regime.
We now show that they actually hold asymptotically in all parts of the
jet.  

Recall that in force-free magnetospheres the enclosed poloidal current
$I$ is preserved along each field
line~\citep{mes61,okamoto1978,tpm86,bes97,nar07,tchekhovskoy_ff_jets_2008},
\begin{equation}
I =  I(\Phi) = \frac{c}{2}\, R B_\phi \approx -\frac{\Omega}{2}\, B_p R^2,
\label{eq_IofPhi}
\end{equation} 
where the approximate equality is due to $B_\phi \approx -E = -\Omega
R B_p/c$ for $R\gg \RLC$ (see eq.~\ref{gammasq}).  Comparing
(\ref{eq_Phi_BpRsq}) and (\ref{eq_IofPhi}) and recalling that $\Omega$
is conserved along field lines, we obtain
\begin{equation}
\Phi \approx -\frac{2\pi}{\Omega}\, I(\Phi).
\end{equation}  
Since both sides of this relation depend only on $\Phi$, this relation
is valid \emph{everywhere} in the solution, even though we derived it
only in the first acceleration regime.  Using~(\ref{eq_IofPhi}) to
substitute for $I$ yields back (\ref{eq_Phi_BpRsq}).  Thus
eqs.~(\ref{eq_Bp_const}) and (\ref{eq_Phi_BpRsq}) are valid everywhere
in the jet.

\section{Coefficients in Equations (52) and (53)}

%(\ref{difeq1}), (\ref{difeq2})

\label{sec_coeffs}

In this Appendix we give explicit expressions for the coefficients
$D_1(R)$--$D_6(R)$ defined in \S3.1.  The functions are
\begin{eqnarray}
D_1(R)&=&-\frac{ck}{\omega}g-\frac{cm}{\omega R}gC_1,
\label{D1} \\
D_2(R)&=&-\frac{cm}{\omega R^2}f-\frac{cm}{\omega R}f'+C_2C_1-\frac{cm}{\omega
R}gC'_1,
\label{D2} \\
D_3(R)&=&\left(\frac{ick^2}{\omega}+\frac{icm^2}{\omega R^2}-\frac{i\omega}{c}
\right)g,
\label{D3}\\
D_4(R)&=&\frac{ck}{\omega}f+\frac{cm}{\omega
R}g-h-C_3\frac{D_3}{D_1},
\label{D4} \\
D_5(R)&=&C_4+C_3\frac{D^2_2}{D^2_1}+\left(C_6+\frac{C_5}{R}\right)C_1
+C_5C'_1-C_7\frac{D_2}{D_1} \nonumber \\
&~& \qquad\qquad\qquad +C_3\left(\frac{D_2D'_1}{D^2_1}-\frac{D'_2}{D_1}
\right)+\frac{ic}{\omega}gC''_1,
\label{D5} \\
D_6(R)&=&C_8+C_3\frac{D_2D_3}{D^2_1}-C_7\frac{D_3}{D_1}+C_3
\left(\frac{D_3D'_1}{D^2_1}-\frac{D'_3}{D_1}\right),
\label{D6}
\end{eqnarray}
where primes denote derivatives with respect to $R$, and the functions
$C_1(R)$--$C_8(R)$ are given by
\begin{eqnarray}
C_1(R) &=& {\omega Rg-cmh\over\omega Rf-ckRh}.
\label{C1} \\
C_2(R)&=&\frac{ck}{\omega R}f-\frac{cm}{\omega
R^2}g-\frac{h}{R}-\frac{h}{R_c}+\frac{ck}{\omega}f'-h',
\label{C2} \\
C_3(R)&=&\frac{ic}{\omega}(f+gC_1),
\label{C3} \\
C_4(R)&=&-\frac{icm^2}{\omega R^2}f+\frac{i\omega}{c}f+\frac{ickm}{\omega
R}g-ikh,
\label{C4} \\
C_5(R)&=&\frac{ic}{\omega R}g+\frac{ic}{\omega R_c}g+\frac{ic}{\omega}g',
\label{C5} \\
C_6(R)&=&\frac{ickm}{\omega R}f-\frac{ick^2}{\omega}g-\frac{2ic}{\omega
R^2}g+\frac{i\omega}{c}g-\frac{im}{R}h,
\label{C6} \\
C_7(R)&=&\frac{2ic}{\omega R}f+\frac{ic}{\omega}f'+C_5C_1+\frac{2ic}{\omega}gC'_1,
\label{C7} \\
C_8(R)&=&\frac{2ck}{\omega R}f-\frac{cm}{\omega
R^2}g-\frac{2h}{R}+\frac{cm}{\omega
R_cR}g-\frac{h}{R_c}+\frac{ck}{\omega}f'+\frac{cm}{\omega
R}g'-h'.
\label{C8}
\end{eqnarray}

%\bibliography{mybib}
%\bibliographystyle{apj}

\end{document}